\documentstyle[11pt,aaspp]{article}

\def\simless{\mathbin{\lower 3pt\hbox
     {$\rlap{\raise 5pt\hbox{$\char'074$}}\mathchar"7218$}}} 

\lefthead{Koerner, Chandler \& Sargent}
\righthead{Dust Disk around DO Tau}

\begin{document}

\title{Aperture Synthesis Imaging\\ of the\\
     Circumstellar Dust Disk\\
     Around DO Tauri}

\bigskip
\author{D. W. Koerner}
\affil{169-506, Jet Propulsion Laboratory,
4800 Oak Grove Dr., Pasadena, CA 91109}
\affil{California Institute of Technology,
Pasadena, CA 91125}

\author{C. J. Chandler}
\affil{National Radio Astronomy Observatory, P.O. Box 0,
Socorro, NM 87801}

\and

\author{A. I. Sargent}
\affil{Division of Physics, Mathematics and Astronomy,
105-24, California Institute of Technology,
Pasadena, CA 91125}

\begin{abstract}
We have detected the T~Tauri star, DO Tauri, in a
0.6$''$-resolution VLA map of 43.3 GHz ($\lambda$ = 7 mm)
continuum emission. The 43 GHz flux density lies on the
same power-law slope defined by 89 to 232 GHz measurements,
F$_\nu$ $\propto \nu^{\alpha}$ with index
$\alpha$ = 2.39$\pm$0.23,
confirming that the 43.3 GHz emission is
thermal radiation from circumstellar dust.
Upper limits to the flux densities
at 8.4 and 22.5 GHz constrain the contribution of free-free
emission from a compact ionized wind to less than 49\%.
The dust emissivity index, $\beta$, is $0.39\pm$0.23, if the emission
is optically thin. Fitting a model of a thin circumstellar
disk to the observed spectral energy distribution gives
$\beta = 0.6\pm0.3$, consistent with the power-law derivation.
Both values are substantially lower than is generally
accepted for the interstellar medium, suggesting grain
growth. Given the youth of DO Tau and the early
evolutionary state of its circumstellar disk, this result implies
that mm-size grains have already formed by the early
T-Tauri phase.

\end{abstract}


%
%
\vfill
\eject

\section{Introduction}

At least 50$\%$ of T Tauri stars (TTs) appear to be surrounded by
circumstellar dust disks (\markcite{Strom et al.\ 1989};
\markcite{Beckwith et al.\ 1990}, hereafter BSCG;
\markcite{Andr\'e \& Montmerle 1994}; \markcite{Henning \& Thamm 1994};
\markcite{Osterloh \& Beckwith 1995}).
Global disk properties can be inferred from models of spectral
energy distributions (SED's) from infrared to millimeter wavelengths
(\markcite{Adams, Lada, \& Shu 1987};
\markcite{Beckwith \& Sargent 1993}; \markcite{Mannings \& Emerson 1994}).
Masses and sizes are
similar to those assumed for the
early solar nebula, suggesting that the disks may be protoplanetary
(cf.\ BSCG; \markcite{Beckwith \& Sargent 1993}).
However, the SED models rely on assumptions about disk morphology
and radial structure, and about the nature of the
constituent dust grains.

Grain size and composition in these potentially
planet-forming disks can be inferred from
the spectral index, $\beta$, of the dust opacity
(cf.\ \markcite{Pollack et al.\ 1994}) if
thermal radiation from grains
in the disk is optically thin. Sub-arcsecond
resolution is necessary to image disks directly and
measure properties on spatial scales of
$\simless$100 AU at the distance of the nearest star-forming regions.
At wavelengths longer than 3mm, the emission
is well into the Rayleigh-Jeans part of the Planck curve and
very likely to be optically thin.
Dust continuum radiation has been detected
from a number of TTs at $\lambda$ =  3~mm
(\markcite{Sargent \& Beckwith 1993} and references therein),
but no thermal emission has been
detected unambiguously at longer wavelengths (Mundy et al.\ 1993).
The required spatial resolution and mJy sensitivities
can now be achieved using the VLA at wavelengths of
7 mm.

DO Tauri is a young TTs in the Taurus star formation complex
at a distance of 140 pc (\markcite{Elias 1978};
\markcite{Kenyon, Dobrzycka, \&
Hartmann 1994}). Estimates of its age and mass
range from 1.6 to 6.0 $\times 10^5$ yrs
and 0.3 to 0.7 $M_{\sun}$ (BSCG;
\markcite{Hartigan, Edwards \& Ghandour 1995}), depending
on the theoretical tracks used to place it on the H-R diagram.
The spectral energy distribution is consistent with the presence of a
$\sim$0.01  $M_{\sun}$ circumstellar disk
(BSCG; \markcite{Beckwith \& Sargent 1991}, hereafter BS;
\markcite{Mannings \& Emerson 1994}).
Asymmetric, blue-shifted, [OI] and [SII]
forbidden line emission
(\markcite{Appenzeller, Jankovics \& \"Ostreicher 1984};
\markcite{Edwards et al.\ 1987}; \markcite{Edwards, Ray \& Mundt 1993})
is resolved as an optical jet at PA 70$^\circ$
(\markcite{Hirth et al.\ 1994}).
The jet is approximately orthogonal to the direction
of linear optical polarization, PA $\sim 170^\circ$
(\markcite{Bastien 1982}), and to
the long axis of CO (2$\to$1) emission
detected in aperture synthesis
images of DO Tau, PA $\sim 160^\circ$ (\markcite{Koerner \& Sargent 1995}).
Kinematic models of the molecular line emission are consistent
with the presence of a circumstellar disk that is
centrifugally supported within a radius of 350 AU from
DO Tau (\markcite{Koerner 1994}).

Here, we report on sub-arcsecond images of the $\lambda$ = 7 mm
emission from DO Tau which were made using the recently
upgraded Very Large Array (VLA) of the National Radio Astronomy
Observatory\footnote{NRAO is operated by
Associated Universities Inc.\ under cooperative
agreement with the National Science Foundation.}.
We have supplemented these measurements with continuum
observations of DO Tau at other wavelengths to sample the spectral
distribution of emission from $\lambda$ = 1.3~mm to 3.6~cm
and improve our understanding of grain properties
in the circumstellar material.

\section{Observations and Results}

The VLA was used to observe DO Tau in radio
continuum emission at 43.3, 22.5, and 8.4 GHz
($\lambda$ = 7 mm, 1.3, and 3.6 cm).
The phase center was offset 1$''$  in both RA and Dec
from the stellar position of DO Tau
(\markcite{Herbig \& Bell 1988}), and the
total bandwidth was 100 MHz in right and left circular
polarizations. As for all observations discussed below,
molecular line emission is negligible within the narrow
band. Observations at 43.3 GHz were carried out on 1994
April 3--4 with the inner seven antennas
of the high-resolution A~configuration,
and on 1994 August 20 with 10 inner antennas of the B configuration.
Baselines up to 5.6 km provided UV coverage in the range
30--800 k$\lambda$.
On both dates, DO Tau was observed at 22.5 and 8.4 GHz
using the remainder of the VLA's
27 antennas. UV coverage was 50--2700 k$\lambda$ at
22.5 GHz and 20--1000k$\lambda$ at 8.4 GHz.
Absolute flux densities were calibrated
using 3C48 and 3C286 with an estimated uncertainty
of 20$\%$. At 43.3 GHz, gain calibration was
accomplished with periodic observations of 0333+321
with a measured flux density of 0.98 $\pm$ 0.06 Jy.
At 22.5 GHz and 8.4 GHz, the gain calibrator was
0400+258 with flux densities 0.65 $\pm$ 0.03 Jy
and 0.83 $\pm$ 0.01 Jy, respectively.

Data calibration and mapping used standard
routines in the NRAO AIPS software package.
Daytime atmospheric phase fluctuations during
A array observations necessitated
extensive editing and application of
a Gaussian taper to the UV data, resulting
in a  $0.68'' \times 0.53''$ (FWHM)
synthesized beam at PA $-78^\circ$.
This corresponds to 95$\times$74 AU at DO Tau.
Fig.\ 1a displays the CLEANed image
of DO Tau at 43.3 GHz. An unresolved source
with flux density 1.80 $\pm 0.71$ mJy is detected at the
stellar position, $\alpha$(1950) = $04^h35^m24.19^s$,
$\delta$(1950) = $26^\circ 04' 54.5''$.
The  $\pm 0.71$ mJy uncertainty includes
rms variations in the map ($\pm$ 0.35 mJy bm$^{-1}$)
and a possible 20\% error in absolute flux calibration.
At 22.5 and 8.6 GHz, DO Tau was not detected
within the area encompassed by the 43.3 GHz
synthesized beam to 3$\sigma$ levels of 0.76 and 0.17 mJy, respectively.

Observations were made with the Owens Valley millimeter array
at 89.2, 111.2, 221.5, and 232.0 GHz
(corresponding to $\lambda$ = 3.4, 2.7, 1.4, and 1.3 mm)
between 1993 September
and 1995 March. Measurements at 89 GHz were made with six
telescopes; four were used at 110 GHz, and five at
220 and 230 GHz. Overall UV-ranges were  5--60 k$\lambda$ (89 GHz),
5--25 k$\lambda$ (110 GHz), and
10--55 k$\lambda$ (220 Hz and 230 GHz).
Resulting FWHM synthesized beams are listed in Table I.
Antenna gains were determined from periodic
observations of 0528+134 and absolute flux density
calibration was based on measurements of Uranus.
Data were calibrated
using the Owens Valley software package, MMA, and mapped with
AIPS.  At all four frequencies,
continuum emission is unresolved
and peaks at the position of the VLA 43.3 GHz image.
Aperture synthesis maps at 89 and 220 GHz are displayed
in Fig.\ 1b and 1c. Measured flux densities at all frequencies
are listed in Table I and displayed in Fig.\ 2.

\section{ Modeling and Discussion}

Our measurements of DO Tau between 8.4 and 230
GHz can be fit by a single power law,
F$_\nu$ $\propto \nu^{\alpha}$,
with index $\alpha$ = 2.39$\pm$0.23,
shown as a solid line in Fig.\ 2.
Earlier detections of radio emission
from TTs at wavelengths greater than 1.3 cm yielded
values of $\alpha$ between 0 and 1
(\markcite{Bieging, Cohen \& Schwartz 1984}) and have been
attributed to free-free
emission from ionized outflows (\markcite{Reynolds 1986}).
Extrapolating from the upper limit of
8.4 GHz emission with $\alpha$ = 1 (dotted line in figure 2),
we estimate that no more than 49$\%$ of DO Tau's 43.3 GHz emission
can arise as free-free radiation from an ionized jet.
Fig.\ 2 suggests the observed mm-wave flux
originates entirely from circumstellar dust.

The frequency dependence
of the mm-wave dust opacity, $\beta$,
can be derived from $\alpha$, since
$\alpha \approx 2 + \beta/(1 + \Delta)$, where $\Delta$
is the ratio of optically thick to optically thin emission from the
disk (BS, eqn.\ 1).
At frequencies where emission from a circumstellar disk
is largely optically thin and the Rayleigh-Jeans approximation
holds, $\beta \approx \alpha - 2$.
For DO Tau, continuum emission appears to be largely
optically thin, even at sub-millimeter wavelengths
(BS; \markcite{Mannings \& Emerson 1994}).
We estimate $0.39 \pm 0.23$ for $\beta_{1-7mm}$, in good agreement
with the BS value of $\beta_{0.6-1mm}$, $0.4 \pm 0.2$.
There is no evidence for the change in $\beta$ longward
of 2mm, postulated by \markcite{Mundy et al.\ (1993)}
for a few other TTs.

An estimate of $\beta$ can also be obtained by fitting the spectral
distribution of luminosity $L_\nu = 4 \pi D^2 \nu F_\nu$ with
a disk model which takes into account any contribution
from optically thick emission.
Following \markcite{BSCG} and \markcite{Adams et al.\ (1990)}, we assumed
power-law radial profiles in disk temperature and surface
density, T = T$_0 (R/R_0)^{-q}$ and $\Sigma = \Sigma_0 (R/R_0)^{-p}$
with p = 1.5 or 1.75.
The millimeter-wave emissivity of the grains, $\kappa$,
is  0.1$(\nu/10^{12} Hz)^\beta$ cm$^2$ g$^{-1}$.
The outer radius, R$_d$,
was allowed to take on values between 22 and 350 AU;
the former is the lower limit to disk size if all 1 mm
emission is optically thick (BS); the latter is the deconvolved half-maximum
radius of the CO-emitting region from aperture synthesis images.
These suggest a disk
inclination angle, $\theta$, of 40$^\circ$
(\markcite{Koerner \& Sargent 1995}).
{}From the 12, 25, and 60 $\mu$m IRAS fluxes, which
probe optically thick regions of the disk, we obtain
T = 218 K at 1 AU with q = 0.54, very close to the value derived
by BSCG for a face-on disk.  Best-fit values of
$\beta$ and M$_d$, the total disk mass, were estimated
from the minimum reduced $\chi^2$ value.
Acceptable fits, with $\chi^2$ falling within
$\Delta\chi^2$ = 1 of its minimum value, were found for our entire
range of p and R$_d$ values. The best-fit model,
with $\beta = 0.6 \pm  0.3$, ${\rm M}_d = 1.0\pm0.5 \times 10^{-2}$
$M_{\sun}$, p = 1.75, R$_d$ = 350 AU, and
$\chi^2$ = 0.77, is plotted in Fig.\ 3 as a solid line,
along with the luminosity distribution derived from IRAS,
sub-millimeter, and millimeter observations of DO Tau.
Following BSCG (eqn.\ 20), these parameters yield
$\Delta \approx 0.28$ at $\lambda$ = 3 mm
and make possible a revised estimate of
$\beta$ from the power-law fit to data presented here.
For $\Delta = 0.28$ and $\alpha = 2.39$, we obtain
$\beta = 0.50 \pm 0.23$, in good agreement with the value obtained
from both our disk-model fit and that of
\markcite{Mannings \& Emerson (1994)}.

For the ISM, it is commonly assumed that $\beta$ is about 2 in the
millimeter wavelength regime (\markcite{Mathis 1990}).
However, a value of 1.3 has been obtained in recent
laboratory studies (\markcite{Agladze et al. 1994}). Even
lower values are suggested by sub-millimeter observations of T Tauri
disks (BS; \markcite{Mannings \& Emerson 1994}). A variety of explanations have
been proposed, including chemical composition,
physical shape, and grain growth
(\markcite{Wright 1987}; BS; \markcite{Kr\"ugel \&
Siebenmorgen 1994}; \markcite{Ossenkopf \& Henning 1994};
\markcite{Pollack et al.\ 1994}).

For DO Tau, we find $\beta \approx$ 0.5 and contend that our denser
sampling of the sub-millimeter regime and spectral coverage extending
to longer wavelengths effectively eliminates uncertainties that may
have complicated other derivations.
Grain properties in circumstellar disks are unlikely to display
the exotic range of chemical composition and physical shapes
required to reproduce this result in the laboratory.
By contrast, the growth of grain size distributions to
include particles larger than 1 mm accounts for our value of $\beta$
(cf.\ \markcite{Miyake and Nakagawa 1993}) and is
consistent with the short theoretical timescales ($\sim 100$ yr)
for production of mm-size particles in the early solar nebula
(cf. Fig.\ 19, \markcite{Cuzzi, Dobrovolskis, \& Champney 1993}).

If the average grain size in disks steadily increases due to
planetesimal formation, $\beta$ should decrease monotonically with age.
However, DO Tau is relatively young, only a
few $\times$ 10$^5$ yrs, with an outflow typical of an active
disk. Grain growth appears to have already occurred by the
early T-Tauri phase.  Recent 7 mm images of a very young protostar,
HH24MMS (\markcite{Chandler et al.\ 1995}), also yield a lower value
of $\beta$ than found for many older TTs in sub-millimeter surveys
(cf.\ BS). These results are inconsistent with a simple
picture of gradually decreasing $\beta$; they could be explained
if mm-size grains grow quickly, followed by
generation of a new population of small dust grains by
planetesimal collisions (\markcite{Lissauer \& Stewart 1993}).
Long-wavelength observations of a statistical sample of
TTs disks encompassing a range of ages are clearly required
to test this hypothesis.

\acknowledgments

We are grateful to D. Wood for assistance during the first season of
43 GHz observations at the VLA.\ \  D.W.K.\ acknowledges support for
this work from NASA grant NGT-51071.
The Owens Valley millimeter-wave array is supported by NSF
grant AST-9314079. Research by A.I.S.\ on protoplanetary disks
is furthered by NASA grant NAGW-4030 from the ``Origins of
Solar Systems'' program. The research described in this
paper was carried out by the Jet Propulsion Laboratory,
California Institute of Technology,
and was sponsored by a fellowship from the National Research
Council and the National Aeronautics and Space Administration.

\begin{table}
\begin{center}
\begin{tabular}{ccccccc}
\tableline
\tableline
Frequency & Flux Density & Statistical & Total &
\hfil Synthesized &
Beam & Parameters \hfil \\
(GHz)  & (mJy) & error (mJy) & error (mJy) & B$_{maj}$
& B$_{min}$ & PA \\
\tableline
\tableline
\phantom{00}8.4 & $<$0.17 (3$\sigma$) & ... & ... & 0.41$''$ & 0.38$''$
& 120$^\circ$ \\
\phantom{0}22.5 & $<$0.76 (3$\sigma$) & ... & ... & 0.25$''$ & 0.23$''$
& $-10^\circ$\\
\phantom{0}43.3 & \phantom{00}1.80 & 0.35 & 0.71 & 0.68$''$ & 0.53$''$
& $-78^\circ$ \\
\phantom{0}89.2 & \phantom{0}14.2\phantom{0} & 0.9 & 3.74 & 2.86$''$
& 2.10$''$ & 82$^\circ$\\
111.2 & \phantom{0}30.2\phantom{0} & 4.5 & 10.5 & 13.0$''$ & 5.4$''$
& $-69^\circ$\\
221.5 & \phantom{0}98.6\phantom{0} & 4.9 & 24.2 & 3.98$''$ & 3.15$''$
& 62$^\circ$\\
232.0 & 137.5\phantom{0} & 4.9 & 32.4 & 3.41$''$ & 3.15$''$ & $-84^\circ$\\
\tableline
\end{tabular}
\end{center}
\bigskip
\tablenum{1}
\caption{Radio (VLA) and mm-wave (OVRO) Continuum Flux Densities from
DO Tauri. Total errors include 20\% uncertainty in the absolute
flux density calibration.}
\end{table}

\clearpage

%
%

%

\clearpage

\end{document}